# The Chandra X-Ray Optics

Martin C. Weisskopf,
NASA, Marshall Space Flight Center

Significant advances in science always take place when the state of the art in instrumentation improves dramatically. NASA's Chandra X-Ray Observatory represents such an advance. Launched in July of 1999, Chandra is an observatory designed to study the x-ray emission from all categories of astronomical objects --- from comets, planets, and normal stars to quasars, galaxies, and clusters of galaxies. At the heart of this observatory is the precision X-Ray optic that has been vital for Chandra's outstanding success and which features an angular resolution improved by an order of magnitude compared to its forerunners. The Chandra mission is now entering its 13-th year of operation. Given that the Observatory was designed for a minimum of 3 years of operation testifies to its robust and carefully thought out design. We review the design and construction of the remarkable telescope, present examples of its usage for astronomy and astrophysics, and speculate upon the future.

## *INTRODUCTION*

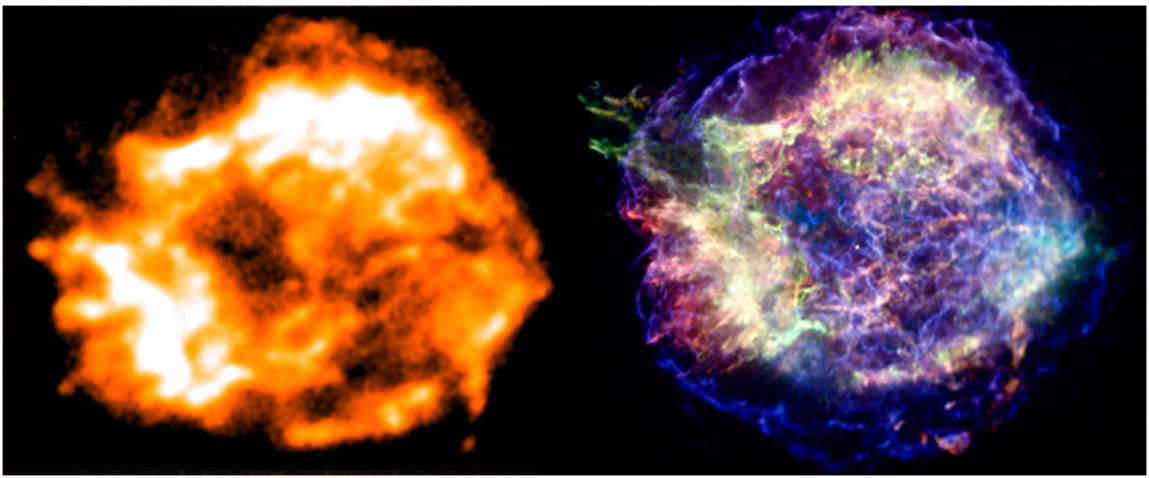

*Figure 1. Two images of the supernova remnant CAS A. Left: the intensity variations seen using the ROSAT satellite. Right: both intensity and spectral variations as viewed with Chandra. The colors on the left represent intensity variations, The colors on the Chandra image represent different energies progressing from 0.5-1.5 keV (red); 1.5-2.5 keV (green); to 4-6 keV (blue). Each image is approximately 7 x 6 arcmin. The sharpness of the Chandra image brings out the point source, the bright white dot at the center of the image to the right, whose presence is only hinted at in the image to the left.[i]*

The Chandra mission has broad scientific objectives and an outstanding capability to accomplish those objectives providing high resolution (< 1-arcsec) images, spectrometric imaging and high resolution dispersive spectroscopy over the energy bandwidth from 0.1 to 10-keV. Figures 1 and 2 exemplify this capability comparing images of astronomical objects taken with Chandra's scientific predecessor, Germany's (with support from the USA and the United Kingdom) Roentgen Satellite (*ROSAT*). Revolutionary for its time (1990-1999), *ROSAT* provided images that pale by comparison to the advancements provided by Chandra.

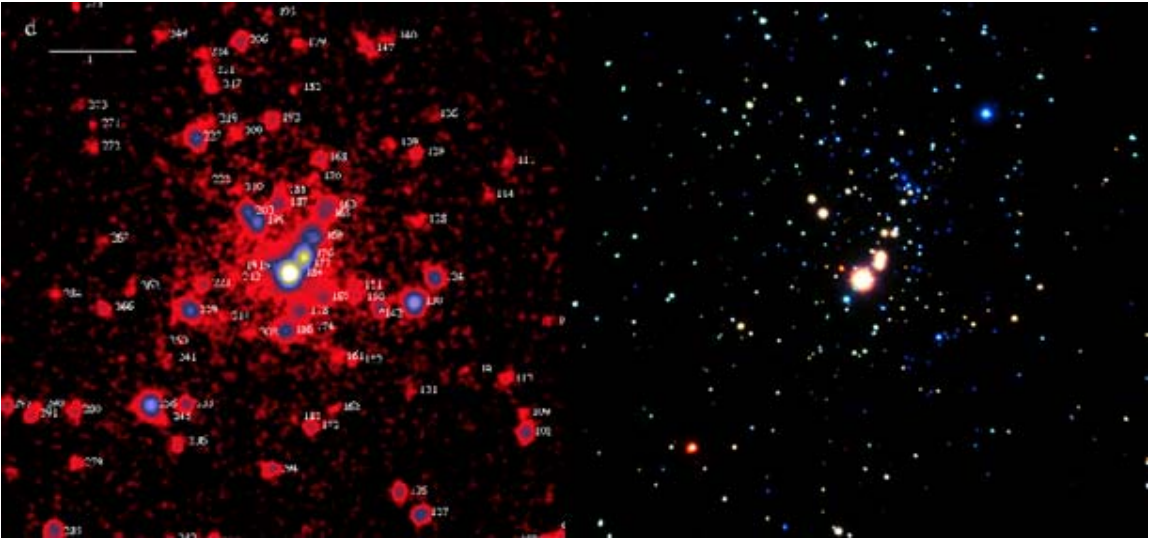

*Figure 2. The X-ray image to the left obtained with ROSAT made the significant discovery of about 75 X-ray sources in the Orion star cluster. The Chandra image to the right resolves about 1500 sources in the same field. Each image is about 5.5 arcmin on a side[ii]. The image is courtesy of NASA at http://chandra.harvard.edu/photo/2005/orion/*

NASA's Marshall Space Flight Center manages the Chandra Project with scientific, technical and operations support from the Chandra X-Ray Center run for NASA by the Smithsonian Astrophysical Observatory in collaboration with the Massachusetts Institute of Technology. TRW's (now NGST) Space and Electronics Group was the prime contractor and continues to provide operations and technical support. Major sub-contracts that related directly to the optics included Hughes Danbury Optical Systems which built the optical elements of the x-ray telescope, Optical Coating Laboratory which coated the optics with sputtered iridium, and then Eastman Kodak Company with responsibility for mounting and aligning the optics and providing the optical bench. Another critical optical system on-board Chandra is a CCD-based, visible-light imaging aspect camera to record star images and provide data for determining (on the ground and after the fact) where the Observatory is pointed to sub-arcsecond precision. Ball Aerospace & Technologies was responsible for the aspect camera system.

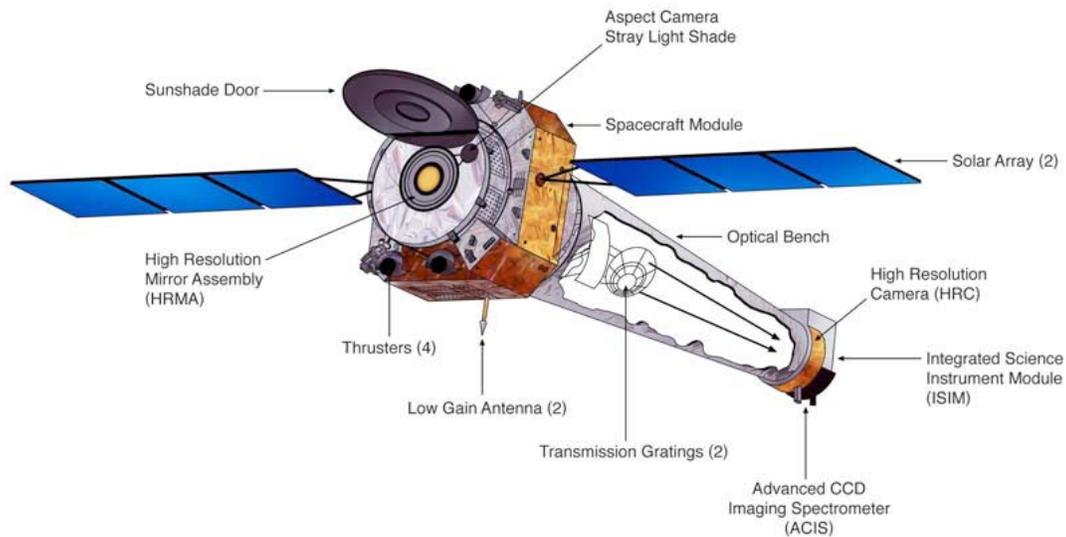

*Figure 3. Schematic representation of the Chandra X-ray Observatory.*

The fully deployed Chandra is shown schematically in Figure 3. The observatory is 13.8 m in length and 19 m in wingspan with a mass of 4,800 kg. The observatory was placed in an elliptical orbit with a nominally 140,000-km apogee and 10,000-km perigee by means of the Space Shuttle Columbia, Boeing's Inertial Upper Stage, and Chandra's own liquid propulsion system. The observatory is comprises three main elements: (1) the telescope containing the X-ray optics, two X-ray transmission gratings[1] that can be inserted into the X-ray path, and a 10-meter long optical bench; (2) a spacecraft that provides electrical power, communications, and attitude control; and (3) a Science Instrument Module (SIM) that holds two focal-plane cameras – the Advanced CCD Imaging Spectrometer (ACIS, Prof. G. Garmire, Pennsylvania State University, Principal Investigator (PI))[1] and the High Resolution Camera (HRC, Dr. S. Murray, Smithsonian Astrophysical Observatory an Johns Hopkins University, PI)[1] – and mechanisms to adjust their position and focus.

ACIS contains two arrays of CCDs that provide position and energy information for each detected X-ray photon. The imaging array is optimized for spectrally resolved, high-resolution imaging over a wide field of view (17 arcminutes); the spectroscopy array, when used in conjunction with the High Energy Transmission Grating (HETG, Prof. C. Canizares, Massachusetts Institute of Technology, PI), provides high-resolution spectroscopy with a resolving power ($E/\Delta E$) up to 1000 over the 0.4-8 keV band. The HRC has two microchannel plate (MCP) detectors, one for wide-field imaging and the other serving as readout for the Low Energy Transmission Grating (LETG, Dr. A. Brinkman, Laboratory for Space Research,, Utrecht, the Netherlands, PI). The HRC detectors provide the highest spatial resolution on Chandra and, in certain operating modes, the fastest time resolution (16 μs). When operated with the HRC's spectral MCP array, the LETG provides spectral resolution >1000 at low (0.08 – 0.2 keV) energies while covering the full Chandra energy band. The Principal Investigators of these instruments are from the Pennsylvania State University, the Massachusetts Institute of Technology, the Smithsonian Astrophysical Observatory and the Space Research Organization of the Netherlands respectively.

---

[1] Readers interested in more detail as to the instruments should consult the Chandra Proposers' Observatory guide and references therein. The Guide may be found at http://asc.harvard.edu/proposer/POG/. See also http://cxc.harvard.edu/cal/, and Weisskopf, et al. 2000, Proc. SPIE Vol. 4012, p. 2-16, X-Ray Optics, Instruments, and Missions III, Joachim E. Truemper; Bernd Aschenbach; Eds, and Weisskopf et al. 2002, The Publications of the Astronomical Society of the Pacific, Volume 114, Issue 791, pp. 1-24.

The system of gyroscopes, reaction wheels, reference lights, and CCD-based star camera enables Chandra to maneuver between targets and point stably while also providing data for accurately determining the sky positions of observed objects. The blurring of images due to pointing uncertainty is <0.10 arcsecond, negligibly affecting the resolution of the optics. Absolute X-ray source positions can typically be determined to ≤0.6 arcsecond[2], providing an unrivaled capability for x-ray source localization. Chandra is currently beginning its 13th year of successful mission operations. The spacecraft and instruments are operating extremely well and have experienced no major anomalies during the mission.

## *The Chandra X-ray Telescope*

### *Description*

The heart of the observatory is the X-ray telescope[3]. The optical design is based on the principal that x-rays reflect efficiently only if the grazing angle between the incident ray and the reflecting surface is less than the critical angle. This angle, typically of order 0.017 radians (one degree), is approximately $10^{-2} (2\rho)^{1/2} /E$, where $\rho$ is the density of the reflecting material in g-cm$^{-3}$ and E is the photon energy in keV. The x-ray optical elements for Chandra resemble shallow angle cones, and two reflections are required to provide good imaging over a useful field of view; the first reflecting surface is a paraboloid of revolution and the second a hyperboloid --- the Wolter 1 configuration. The collecting area is increased by nesting concentric mirror pairs, all having the same focal plane.

The optical elements have four paraboloid-hyperboloid pairs with a common ten meter focal length. The element lengths are about 0.83-m, the diameters approximately 0.63, 0.85, 0.97, and 1.2-m, and wall thickness range from 16-mm for the smaller elements to 24-mm for the outer ones. Zerodur from Schott is the optical element material chosen because of its low coefficient of thermal expansion and demonstrated capability of permitting very smooth polished surfaces.

---

[2] Approaching 0.1" if known optical or infrared sources in the field of view are detected as X-ray sources.
[3] Readers interested in more details as to both the Chandra optics and X-ray optics in general should consult Chapter 4 of the Chandra Proposers' Observatory Guide and the references therein. The Guide may be found at http://asc.harvard.edu/proposer/POG/.

*Fabrication*

The major fabrication phases included coarse and fine grinding, polishing, and final smoothing. The grinding and polishing operations were done with relatively small tools under computer control. Cycles were iterative; a mirror element would be measured to yield an error map, appropriate tools selected to reduce the errors, and a polishing control file for the next cycle generated. The next cycle would cause more material removal in the high areas. The residual errors would be smaller than previously, and so the process converged to the required accuracies.

Three primary metrology instruments were used. Axial figure errors at fixed azimuthal angles were measured in a Precision Metrology Station (PMS) by interferometrically measuring the separation between the optical surface and a calibrated reference. A Circularity and Inner Diameter Station (CIDS) was used to determine inner diameters and roundness errors near the ends of the reflecting elements; these instruments each included two pairs of opposed contacting radial probes, calibrated reference bars, positioning mechanisms, and a precise air bearing to permit rotation. The instruments were housed in environmentally controlled enclosures, and the mirror elements were carefully supported in the vertical orientation to achieve the required accuracy. Finally, the Micro-Phase Measuring Interferometer (MPMI), an adaptation of a WYKO non-contact profilometer, was used to obtain high-frequency roughness measurements of small samples of the surfaces. In general, x-ray performance is more sensitive to high-frequency than to low-frequency errors, and to axial than to circumferential errors. The point spread function (PSF) for an X-ray optic may be viewed as consisting of two components, one resulting from these high-frequency errors (surface micro-roughness) which scatters photons out of the central core of the PSF, and this core whose properties are determined by larger scale figure errors. The sensitivity to these errors was reflected in the accuracies of the instruments, about 0.5, 20, and 200-Å-rms for the MPMI, PMS, and CIDS respectively. Numerous cross-checks, including comparisons with data from other (but less accurate) instruments and different optical orientations, were performed during fabrication to avoid serious systematic errors.

The PMS axial data and CIDS circumferential data combined yielded low frequency surface error maps and these errors were then reduced to an average of 50-Å-rms for the axial component by computer controlled polishing with small tools. This typically required four grinding and four polishing cycles per element. Figure 4 shows the one of the optical elements during this process. The aluminum rings visible in the figure provided support to the optics during the abrasive material removal process, and were removed during (vertical) measurements so that an essentially stress-free shape could be determined; heater elements on the rings were used to effect glass insertion or removal by differential expansion of the glass and support structure. Final polishing was performed with a large lap designed to reduce surface roughness without introducing unacceptable lower frequency figure errors. The resulting rms surface roughness over the central 90% of the elements varied between 1.85 and 3.44Å in the 1 to 1000-mm$^{-1}$ band; this excellent surface smoothness improved the encircled energy performance at higher energies over initial specifications..

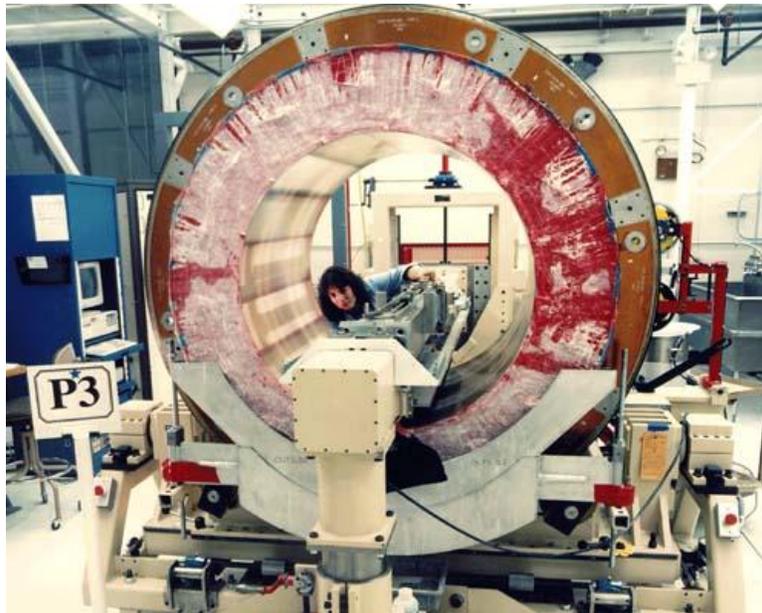
*Figure 4. One of the paraboloids during rough grinding*

*Coating*

The mirror elements were coated at Optical Coating Laboratories, Inc. (OCLI) by sputtering with iridium over a binding layer of chromium. Before each coating optical witness samples were used to show that the thickness would be uniform and that the surface smoothness was not degraded. The x-ray reflectivity of these witness flats was also measured to verify coating density. Witness samples coated simultaneously with the flight elements were also tested. The final cleaning occurred at OCLI prior to coating, and subsequently stringent contamination controls were put into place to minimize particulate and molecular contamination.

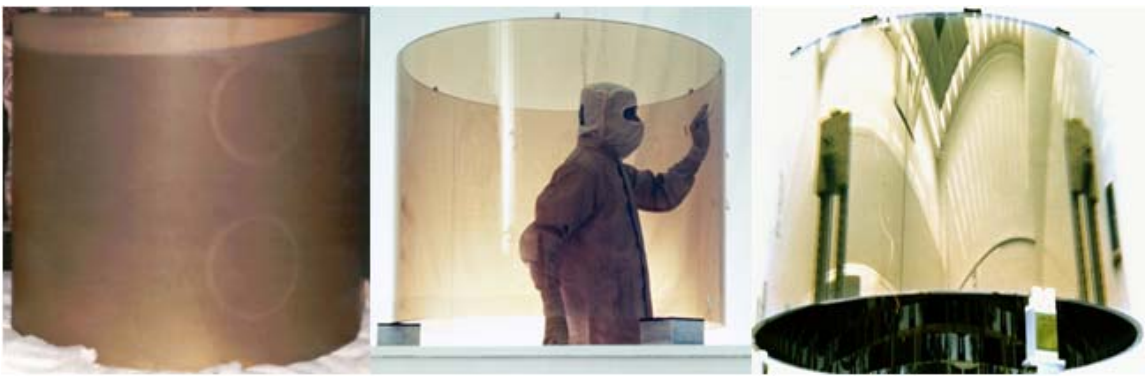

*Figure 5. Chandra telescope elements at different stages of production. Left: a single raw blank prior to grinding and polishing. Center: polished element. Right: coated element prior to insertion into the mirror holding fixture.*

*Assembly and Alignment*

The final alignment and assembly of the mirror elements into a telescope was done by the Eastman Kodak Company (EKC). The mirror element support structure, prior to inserting the reflecting elements, is shown in Figure 6. Each mirror element was bonded near its middle to flexures attached to carbon fiber composite support sleeves. The four support sleeves and associated flexures for the paraboloids can be seen near the top of the figure. The flexures produce only very small radial forces on the mirrors, and therefore reduce support-induced axial slope errors. The thin mirror shells are also susceptible to a deformation mode in which both ends become oval, but with perpendicular major axes. Supporting the mirror elements near their centers minimized the coupling of support errors into this mode.

The final assembly and alignment was performed with the optical axis vertical in a clean and environmentally controlled tower. The mirror elements were supported to approximate the gravity- and strain-free state, positioned mechanically and optically, and then bonded to the flexures. A Bauer Associates optical instrument was used for alignment and generated a laser beam which passed through the x-ray optics, reflected from an auto-collimating flat, and returned through the x-ray optics to the instrument. The variation of the returned spot position with azimuth provided the data necessary to position the elements.

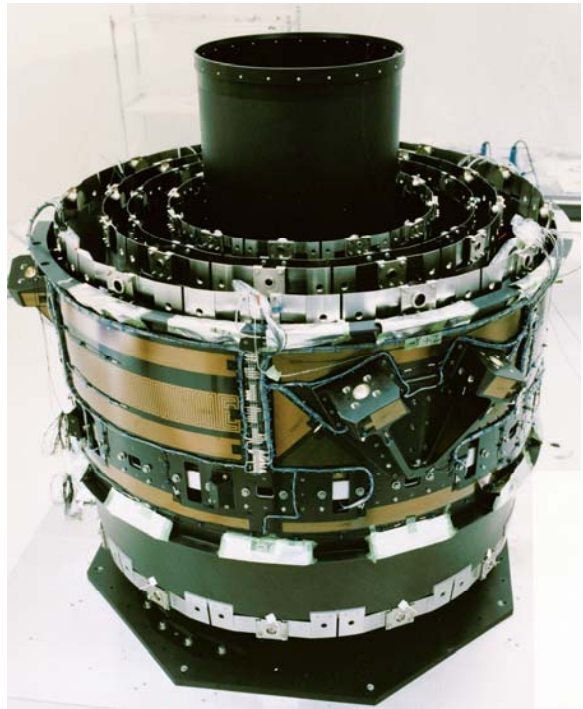

*Figure 6. Chandra telescope mirror support structure.*

*Calibration*

The telescope was taken to NASA's Marshall Space Flight Center for end-to-end ground x-ray calibration beginning in December 1996. The calibration was performed over a 6 month period at the Center's X-ray Calibration Facility (Figure 7). Previously, the largest parabolod-hyperboloid pair, uncoated and uncut to its final length, had been x-ray tested at the Facility during an earlier phase of the development program. This early test showed a measured angular resolution of 0.22 arcsec (FWHM). During the activities in 1996-1997 all the flight instrumentation (ACIS, HRC, the LETG and HETG) were tested with the telescope. Moreover, additional calibrations were performed using the telescope and synchrotron-facility-calibrated non-flight detectors. The on-ground calibration results demonstrated, in advance of launch, that the Chandra X-Ray Observatory would provide the required science capabilities: high-resolution (sub-arcsec) imaging, high resolution spectrometric imaging, and high-resolution dispersive spectroscopy.

The ground calibration of Chandra was the most comprehensive such calibration for an X-ray mission ever performed and this remains true to this date. The intensive calibration served many useful purposes above and beyond simply verifying functional performance. Additional benefits included such items as accelerating and testing the development of the scientific data system to team-building as a result of the 24/7 environment. Note that the Facility is now being used for testing and calibration of the James Webb Space Telescope optics, a function that could not have been conceived of at the time of the Facility's construction and a demonstration why building national test facilities have unanticipated benefits.

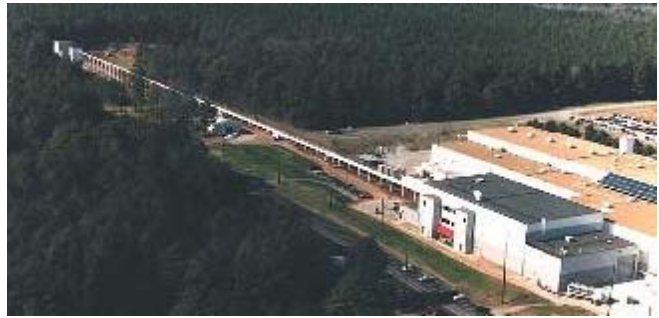

*Figure 7. The X-Ray Calibration Facility at NASA's Marshall Space Flight Center. The small building to the far left houses the X-ray Source System including x-ray generators, monochrometers, and filters; the large building to the near right houses the instrument chamber. The distance from the XSS to the location of the entrace aperture of the telescope in the instrument chamber was 525-m facilitating simulating point sources.*

### Science with CHANDRA

Chandra's unique ability to obtain sub-arcsecond X-ray images and its unrivaled sensitivity for detecting faint sources and measuring high-resolution X-ray spectra establish it as an integral contributor to the progress in astrophysics today and for the foreseeable future. Chandra researchers have investigated phenomena as diverse as the spectra of Jupiter's auroras and the effect of dark energy on the growth of galaxy clusters. Chandra has observed X-ray sources with fluxes spanning 10 orders of magnitude from the first-discovered and x-ray-brightest Sco X-1 to the faintest sources in deep fields. Given the superb capabilities of the optics and associated instrumentation, the scientific possibilities are enormous. Some of the most exciting investigations have resulted from unexpected discoveries that improved sensitivity produces. Chandra provides better angular resolution in the X-ray band by about a factor of 10, and much improved and more efficient spectroscopic resolution than previous x-ray astronomical observatories. In the following we describe but a handful of the exciting discoveries made with Chandra over the past 12 years.

*Normal Stars*

The vision that normal stars would be detectable as x-ray sources came decades ago with the knowledge that our sun possessed an "outer skin" the solar corona - at very high, x-ray-emitting temperatures of several million degrees. The forerunners of Chandra, especially the *EINSTEIN* Observatory and *ROSAT,* proved this conjecture correct. Now x-ray emission has been detected using Chandra from practically all classes of normal stars. Observations such as shown in Figure 2 have now provided a major advance in the capability to measure coronal plasma parameters --- knowledge of which are critically needed to extend our understanding of stellar coronae in general and, by comparisons, to our own sun in particular.

*Supermassive Black Holes*

Amongst the many exciting results enabled by Chandra is the discovery that there appears to be a cycle of feedback occurring between supermassive black holes and their surroundings. Oversimplifying, the energy of jets of particles emanating near the black hole spews out into the surrounding medium, which, in the case of supermassive black holes in active galaxies (themselves in clusters of galaxies), can influence the behavior of the entire cluster. The energy carried off by the jets clears out the surrounding medium which cannot then feed material to the black hole. Ultimately the energy carried off in the jets dissipates now allowing material to once more to feed the black hole. Recently Chandra has recently discovered evidence for this feedback mechanism at work on much smaller scales. Chandra HETG observations of the circumnuclear environment of an individual galaxy (the Seyfert galaxy NGC 1068, Figure 8)[iii] provides data which shows that a large amount of material (more than the equivalent of one sun per year) is deposited into the interstellar medium of this galaxy. Such high resolution spectroscopy is only possible because of the optics.

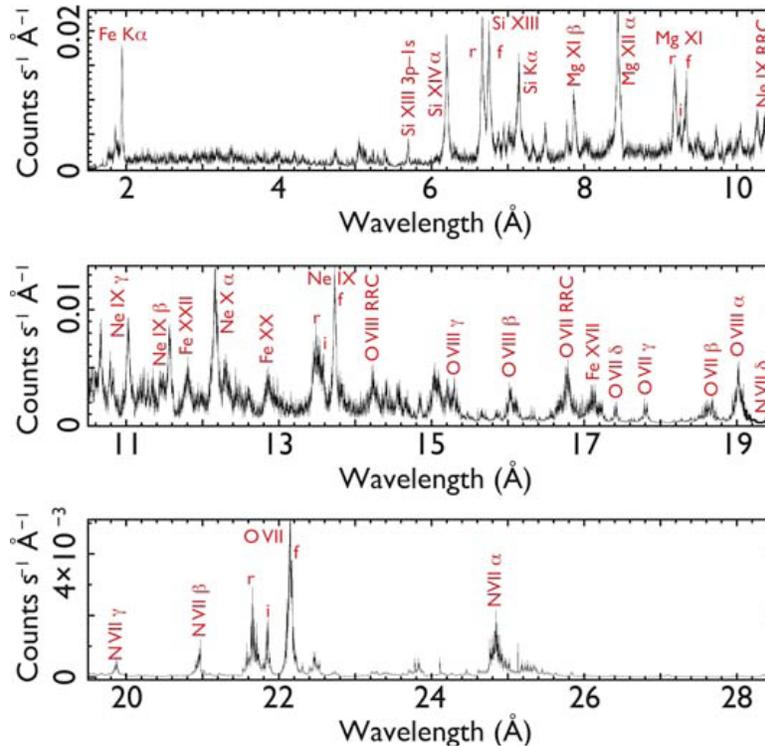

**Figure 8. Chandra HETG spectrum of the nucleus of NGC 1068 with Hydrogen-like and Helium-like transitions including resonance (r), intercombination (i), forbidden components (f), and radiative recombination continua (RRC) labeled.** [iii]

Astronomers obtained an image of what is known as the Chandra Deep Field South (CDFS) by pointing the telescope at the same patch of sky for over ten weeks of time. Figure 9 shows small section of the image with HST data superimposed. The Chandra data allowed astronomers the most sensitive search for black holes in 200 optically-identified distant galaxies, from when the Universe was between about 800 million and 950 million years old. These galaxies were so faint in X-rays that none of the galaxies was individually detected, but because of accurate position determinations by Chandra, one could add up all the X-ray counts near the positions of these distant galaxies. The result of this work is somewhat controversial[iv,v,vi,vii] and it appears that no such X-ray galaxies were detected. Despite firm detections these results shed further light on the relationship between galaxies and the supermassive black holes.

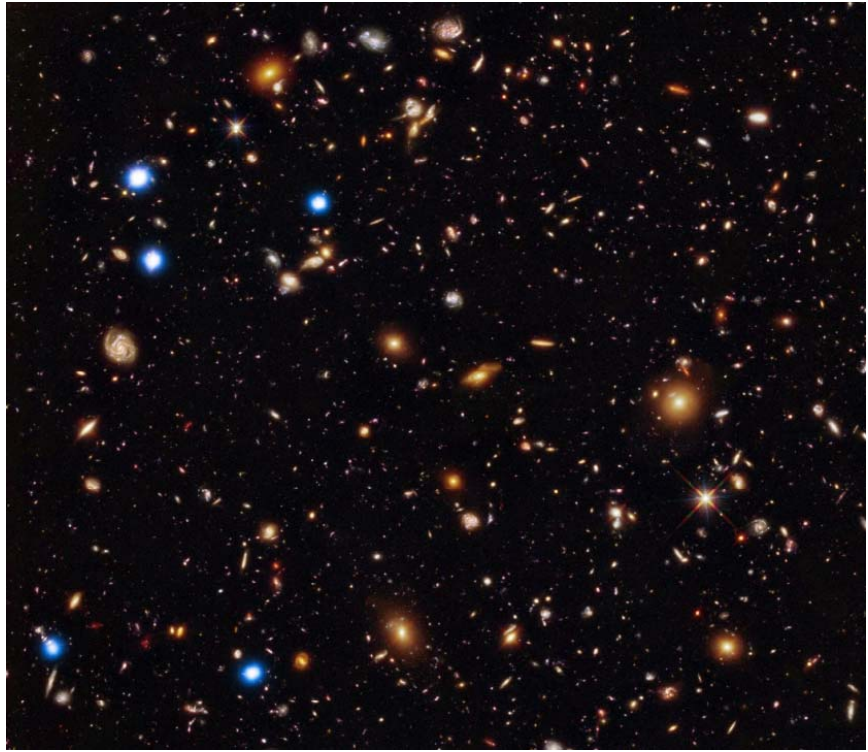

*Figure 9. This composite image from Chandra (blue sources) and the Hubble Space Telescope (yellow)) combines the deepest X-ray, optical and infrared views of the sky. The image is a subsection of the Chandra Deep Field South. The image is approximately 24-arcseonds on a side.[iv] The image is courtesy of NASA at http://chandra.harvard.edu/photo/2011/cdfs/ .*

*Clusters of Galaxies: Dark Matter, the Hubble Constant and Dark Energy*

The determination and location of mass in the Universe is a principal problem of current interest. Observations with Chandra, working in conjunction with optical telescopes, have provided perhaps the most dramatic evidence for the presence of the so-called "dark matter" (dark because we don't see it - yet we know that it is there) - on a variety of astronomical spatial scales including galaxies and clusters of galaxies. It is now well known e.g., that the total mass in galaxy clusters is larger than the mass inferred from the luminous stars in the clusters. We also know, from previous x-ray observations, that the clusters are filled with an extremely hot x-ray emitting gas. The mass in the gas albeit about an order of magnitude more than may be found in the stars in the galaxies, does not account for the total mass that clearly holds the entire cluster together through gravity. But the hot x-ray-emitting gas does serve as an ideal tracer to map the mass (really the gravitational potential). Chandra has proven itself ideal for this type of research. A significant milestone in the investigation of dark matter was the use of a combination of Chandra and optical observations where the effect of the total mass of the clusters is measured by the gravitational lensing of more distant astronomical objects.

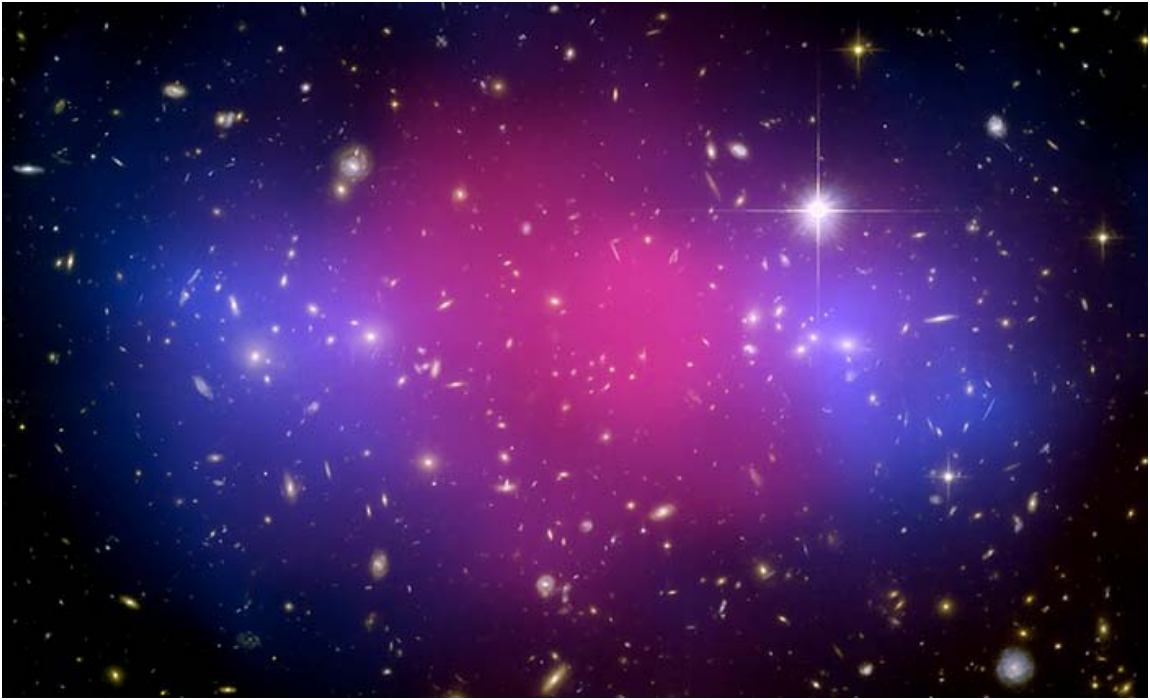

*Figure 10. Composite image of MACS J0025.4-122, an object produced by a merger between two galaxy clusters with similar mass at z=0.586, shows hot gas (pink) detected by Chandra and location of most of the cluster mass (blue) from gravitational lensing measurements using HST and Keck. The image is 3.2 arc-minutes across.[viii] The image is courtesy of NASA at http://chandra.harvard.edu/photo/2008/macs/.*

Results such as those in Figure 10 demonstrate the separation of the dark matter from the X-ray gas mass as a result of a collision between two clusters of galaxies. This image not only directly shows that the mass in clusters is predominantly in some dark form that interacts with baryonic matter only through gravity, but also that the dark matter does not strongly interact with itself --- an important finding to help lead one to the understanding of the properties of dark matter.

Observations of the hot gas in galaxy clusters may also be used, in conjunction with ground-based radio measurements, to determine the distance to clusters and hence the Hubble constant. The approach relates observations of the x-ray-emitting gas with Chandra and the shift in frequency of the 3 degree (K) microwave background photons (the relic of the big bang) occurring during scattering off the energetic, x-ray emitting electrons (the Sunyaev-Zel'dovich effect). These two measurements, when combined, determine the distance to the cluster and hence the Hubble constant, in a relatively model independent fashion and over much larger distances than accessible to more traditional techniques. The derived value of 77 km s$^{-1}$ Mpc$^{-1}$ is consistent with the more local measures.[ix]

Finally, observations with Chandra as to the rate of growth of clusters of galaxies versus time provide not only evidence for the effects of dark energy on the growth of structure in the universe, but also important and independent constraints on cosmological parameters. Figure 11 shows the results of applying the Chandra studies of this rate of growth to constrain certain cosmological parameters: the equation of state of the universe and the energy density of the dark energy that is accelerating the universe apart.

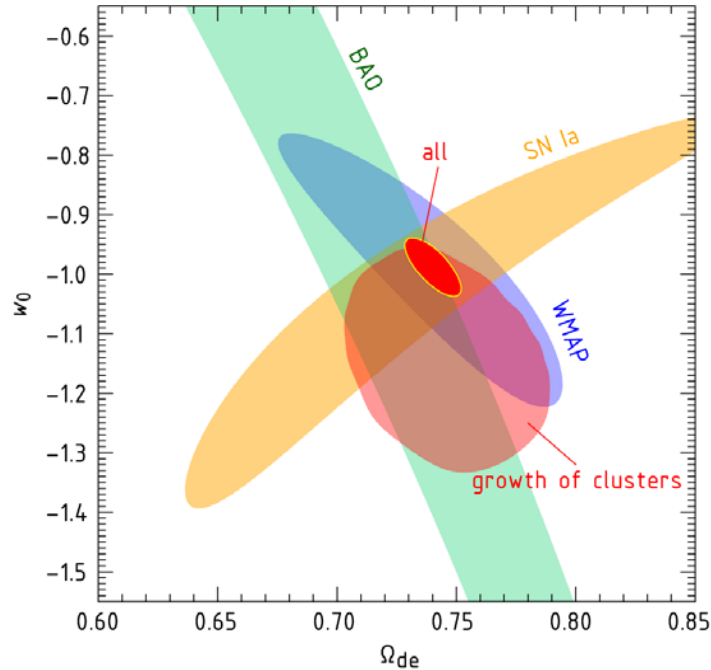

*Figure 11. Plot of the cosmological parameters $w_0$ (the equation of state) as a function of the energy density of dark energy as constrained by different observations: "growth of clusters" is derived from Chandra measurements of the rate of growth of the X-ray emitting clusters of galaxies as a function of cosmic time, "SN Ia" are constraints produced by optical measurements of the apparent intensity of supernovae type Ia, WMAP refers to constrains deduced from the Wilkenson Microwave Anisotropy Probe which deduce constraints based on structure in the microwave background; "BAO" refers to constraints derived from optical measurements of Baryon Acoustic Oscillations.[x]*

## *Conclusions*

The Chandra X-Ray Observatory, from its launch to today, continues to have a profound influence on astronomy and astrophysics. The data from the Observatory are used by scientists throughout the world. Demand for its use is very high with the number of proposals submitted and the request for observing time significantly exceeding what is available. The Chandra program is extremely highly ranked when compared to other space-based and ground based observatories as evidenced by its grades after NASA's Senior Reviews, by independent studies of its impact on astronomy and astrophysics through publications and citations to those publications, by the numbers of PhD theses produced, etc. Moreover, especially considering that Chandra's unique capabilities for angular resolution that will not be approached by any projected new missions throughout the world for at least the next decade, Chandra is scientifically necessary to play its role for the foreseeable future.

*Acknowledgements*

As Project Scientist for the Chandra X-Ray Observatory since 1977, I have authored or co-authored numerous papers and documents describing all aspects of the Chandra Project. This paper is a synthesis of several of these papers and documents. Thus, I gratefully acknowledge the help and support of the many scientists that have, contributed in so many ways to the Chandra Project. I especially want to acknowledge Leon Van Speybroeck, the Telescope scientist, without whose efforts the optics would not have performed so well, Harvey Tananbaum, the Director of the Chandra X-Ray Center, without whose efforts there would be no observatory, Stephen L. O'Dell, my Deputy, who has made my job so much easier, and Riccardo Giacconi for the vision and insight that inspired us all.